\documentclass[prd,twocolumn,amssymb,amsmath,nofootinbib,longbibliography]{revtex4-1}

\usepackage{amssymb,amsmath,mathrsfs,epstopdf,slashed,color,float}
\usepackage{dcolumn}   % needed for some tables
\usepackage{bm} 
\usepackage{tikz}
\usetikzlibrary{matrix}
\usetikzlibrary{positioning}
\usepackage{ifpdf}
\ifpdf
\usepackage{graphicx}
\usepackage{hyperref}
\else
\usepackage[dvipdfmx]{graphicx}
\usepackage[dvipdfmx]{hyperref}
\usepackage[numbers]{natbib}
\usepackage{hypernat}
\fi
\DeclareMathOperator*{\Tr}{Tr}

\newcommand\ba{\begin{eqnarray}}
\newcommand\ea{\end{eqnarray}}
\newcommand\be{\begin{equation}}
\newcommand\ee{\end{equation}}

\def\beq{\begin{equation}}
\def\eeq{\end{equation}}
\def\rmd{\mathrm{d}}
\def\t{\times}

\begin{document}

\title{Role of Bloch waves in baryon-number-violating processes}  

\author{Yu-Cheng Qiu$^{1}$}
\email{yqiuai@connect.ust.hk}
\author{S.-H. Henry Tye$^{1,2}$}
\email{iastye@ust.hk, sht5@cornell.edu}
\affiliation{$^1$ Department of Physics and Jockey Club Institute for Advanced Study, \\Hong Kong University of Science and Technology, Hong Kong SAR 999077, China}
\affiliation{$^2$ Department of Physics, Cornell University, Ithaca, New York 14853, USA}

\date{\today}
  
\begin{abstract}
In the Bloch-wave approach to estimate the baryon-number-violating scattering cross section in the standard electroweak theory in the laboratory, we clarify the relation between the single sphaleron barrier and multiple (near periodic) sphaleron barrier cases. We explain how a realistic consideration modifies/corrects the idealized Bloch wave and the resonant tunneling approximation. The basic approach is in part analogous to the well-known triple-$\alpha$ process to form carbon in nucleosynthesis.
\end{abstract}

\maketitle

\section{Introduction}

The $SU(2) \times U(1)$ electroweak (EW) theory is very well established by now. With the $SU(2)$ gauge coupling $g\simeq 0.645$ (or $\alpha_W=g^2/4\pi \simeq 1/30$), the W-boson mass $m_W=gv/2\simeq 80$ GeV (where $v=246$ GeV is the Higgs vacuum expectation value), and the Higgs mass $m_H=125$ GeV all measured, the theory (without further extension) has no free parameters and all dynamics are in principle completely determined. One important property is the sphaleron potential barrier height, also known as the sphaleron mass/energy, $E_\text{sph}=9.0$ TeV [turning off the $U(1)$ coupling raises the mass to 9.1 TeV], which separates vacua with different values of the Chern-Simons number $n$~\cite{Manton:1983nd,Klinkhamer:1984di}. 

Due to the presence of instantons and anomalies, the baryon number $B$ and lepton number $L$ are not conserved in EW theory~\cite{tHooft:1976rip,tHooft:1976snw}. Thus, one would like to search for these ($B+L$)-violating processes in the laboratory, where the changes $\Delta B=\Delta L= 3\Delta n$ and $\Delta n$ is the change in $n$. Interesting parton (left-handed quarks) scatterings in proton-proton collisions include $\Delta n \ne 0$ scatterings at quark-quark energy $E_{qq} \ge E_\text{sph}$; e.g., 
 a $\Delta n =+1$ quark-quark scattering,
\begin{equation}\label{qqBL+1}
u_L + u_L \rightarrow e^-\mu^- \tau^- bbbcccuuuuu+ X
\end{equation}
where $X$ includes particles that conserve electric charge as well as ($B-L$). A single ($B+L$)-violating event can produce three negatively charged leptons plus three $b$ quarks (a $b$ quark can be replaced by a $t$ quark). Other interesting possible experimental detections have also been proposed recently~\cite{Ellis:2016ast,Brooijmans:2016lfv,Ellis:2016dgb}. 

\begin{figure}
	\centering
	\includegraphics[width=8.8cm]{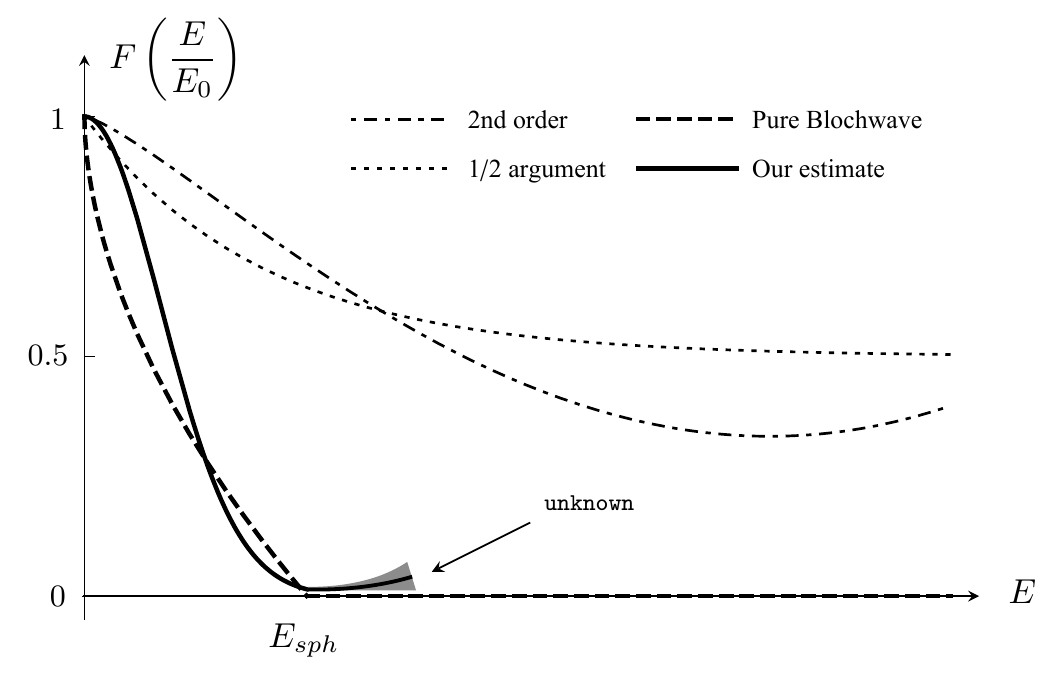}
	\caption{Schematic estimate of the function $F({E}/{E_{0}})$. The dot-dashed curve follows the expansion up to $\mathcal{O}((E/E_{0})^2)$~\cite{Ringwald:1989ee,Espinosa:1989qn,Khoze:1990bm,Mueller:1991fa,Khlebnikov:1990ue}. The dotted curve \cite{Bezrukov:2003yf} reaches $1/2$ as $E \to \infty$. The dashed curve is for the idealized Bloch wave analysis \cite{Tye:2015tva}. The solid curve is the best estimate of this paper.}
	\label{FE}
\end{figure}

Although it is well known that baryon-number-violating processes happen in EW theory~\cite{tHooft:1976rip,tHooft:1976snw,Manton:1983nd,Klinkhamer:1984di}, there is a large ($\sim 70$ orders of magnitude) discrepancy in the determinations of the baryon-number-violating scattering cross sections at $E \sim E_\text{sph}$~\cite{Ringwald:1989ee,Espinosa:1989qn,Khoze:1990bm,Mueller:1991fa,Zakharov:1990dj,Mattis:1991bj,Rubakov:1996vz,Ringwald:2002sw,Ringwald:2003ns,Bonini:1999kj,Bezrukov:2003yf,Tye:2015tva,Tye:2017hfv}, separating the observable from the totally unobservable predictions in the laboratory. Early estimates showed that the ($B+L$)-violating scattering cross section in the laboratory goes like
\begin{align}
\label{Espin}
\sigma (\Delta n \ne 0) \propto & \exp {\left(-\frac{4\pi}{\alpha_W} F(E/E_{0})\right)} 
\overset{E\to 0}{\sim} 10^{-164}
\end{align}
where $E_0=\sqrt{6}\pi m_W/\alpha_W\simeq 18.5$ TeV and $F(E=0)=1$. Leading-order corrections show that $F(E)$ decreases ($\sigma$ increases) as $E$ increases, but the estimate is no longer reliable for $E \to E_\text{sph}$. Although it is believed that $F(E) \ne 0$ for any energy, one cannot rule out the possibility that $F(E)$ becomes small enough at $E \gtrsim E_\text{sph}$ so that the exponential factor is no longer suppressive. However, earlier speculations have argued that such a ($B+L$)-violating scattering cross-section $\sigma(E_{qq},{\Delta n} \ne 0)$ in the laboratory remains exponentially small (see Fig.~\ref{FE}); even if one could reach proton-proton energies of around 50 TeV, with the quark-quark energy $E_{qq}$ much higher than the sphaleron barrier height of 9 TeV, the event rate would still be far too small to be observed~\cite{Rubakov:1996vz,Bezrukov:2003er,Bezrukov:2003qm,Ringwald:2003ns,Ringwald:2002sw}. 

It is useful to take an entirely different approach to this problem. In the idealized situation, one starts with the Bloch-wave formulation for the periodic sphaleron potential~\cite{Tye:2015tva}. Because of the parton distribution function, $E_{qq}$ has an energy spread. For energy $E_{||}$ along the ($B+L$)-violating direction within a conducting (passing) band, the ($B+L$)-violating process is unsuppressed. Since there is no solution for $E_{||}$ outside the band, only the ($B+L$)-conserving process can take place for $E$ in the Bloch-wave band gap. At $E \sim 0$, due to the very narrow Bloch-wave band width~\cite{Tye:2015tva},
\begin{align}
\label{BTS}
 \sigma (\Delta n \ne 0) \propto  \frac{\rm band width}{\rm band gap + band width} 
 \sim  10^{-179}
 \end{align}
Up to the (different) prefactors, \eqref{BTS} qualitatively reproduces the exponential suppression \eqref{Espin} from a totally different viewpoint. The advantage of this viewpoint is the reliability of the extrapolation to higher energies. In this approach, we find that $\sigma (\Delta n \ne 0)$ is no longer exponentially suppressed at $E \sim E_\text{sph}$, so there is a chance that a ($B+L$)-violating event may be observed at the Large Hadron Collider (LHC). The discrepancy is illustrated in Fig.~\ref{FE}.

Reference~\cite{Tye:2017hfv} briefly compared these two very different estimates of the ($B+L$)-violating scattering cross section. Here, we shall provide some background and clarifying discussions on the Bloch-wave analysis~\cite{Tye:2015tva}. However, the idealized Bloch-wave picture also has its own corrections/modifications, including the following: \\
1) The presence of baryon-number-conserving directions, which allows for a leaking of energy (or ``decay") from the ($B+L$)-violating direction, indicated by a drop in $E_{||}$ which is a transition from a higher energy band to a lower energy band, where $E_{||}$ is the energy along the the ($B+L$)-violating direction while $E_{qq}$ ($\ge E_{||}$) includes energies along the the ($B+L$)-conserving direction.\\
2) The existence of fermion masses effectively tilts the periodic sphaleron potential, as illustrated in Fig.~\ref{p}. This tilt of the potential also assures us that its translational symmetry is global instead of ``local."

In the absence of both the ``decay" and tilt of the periodic potential, which is the idealized Bloch-wave case, $\Delta n$ is unbounded. In the presence of tilt but no ``decay," $\Delta n =0$, since the incoming wave would eventually hit the totally inaccessible region (as shown in Fig.~\ref{p}) and be totally reflected. In the realistic situation where both decay and tilt are present, we argue that $\Delta n$ is finite and small, probably dominated by $\Delta n=\pm 1$.

To have an unsuppressed ($B+L$)-violating event rate, we need energy along the periodic sphaleron potential direction $E_{||} \sim E_\text{sph}$, which means that the quark-quark energy $E_{qq} \gtrsim E_\text{sph}$. At proton-proton energy $E_{pp}=14$ TeV, the parton distribution suppression is $\sim 10^{-6}$. Together with the phase-space suppression [$(1-\sqrt{E_{||}/E_{qq}})^2 \sim 10^{-4} \sim 10^{-3}$~\cite{Tye:2017hfv}], this leads to an overall suppression of $\sim 10^{-10}$. A crude estimate suggests that there may be $10^3$ observable ($B+L$)-violating events during the HE-LHC run. (The suppression is a few orders of magnitude worse at $E_{pp}=13$ TeV.) Such a suppression will be substantially alleviated if the LHC can increase $E_{pp}$ just a few TeV above 14 TeV.

The rest of this paper is organized as follows. Section~\ref{issue} reviews the issue of the rate of the ($B+L$)-violating process.  Section~\ref{BlochR} reviews the basic physics, with an emphasis on the periodic potential that we address here. In Appendix~\ref{bands} we show the existence of continuous (in energy) Bloch-wave bands. Here, we emphasize that EW theory has Bloch-wave bands but no $\theta$ vacuum, in contrast to QCD, which has the $\theta$ vacuum but no Bloch-wave bands. Section~\ref{ResD} gives a general description of resonant tunneling with decay. Section~\ref{single} discusses the single barrier with decay model, which already captures many of the key features. This is followed by discussions for the double barrier case in Sec.~\ref{Double} and the triple barrier case in Sec.~\ref{Triple}, where the tilt is included. These cases allow us to see the general features of a tilted periodic potential with decay, as discussed in Sec.~\ref{multiple}. Section~\ref{Disc} gives a brief discussion. Appendix~\ref{triple} gives a brief review of the famous triple-$\alpha$ process, a prime example of resonant enhancement through tunneling and decay. Appendix~\ref{Mink} reviews the sphaleron process, in particular the shape of the sphaleron in Minkowski spacetime.

\section{The Issue}
\label{issue}

\begin{figure}
\centering
\includegraphics[height=5cm]{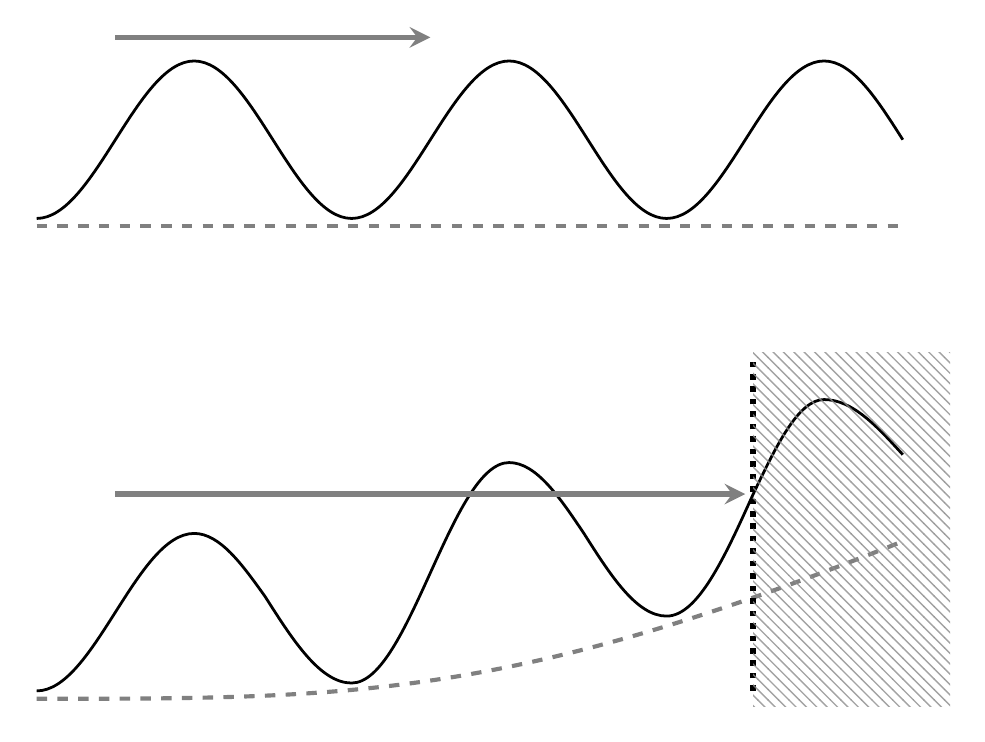}
\caption{The upper graph shows the idealized quantum-mechanical periodic sphaleron potential as a function of the Chern-Simons variable. The barrier height is $E_\text{sph}$. It has Bloch waves as solutions. The lower graph shows the actual sphaleron potential relevant in the presence of fermions, where the tilt is exaggerated for illustrative purposes. The shaded area implies that the inaccessible region will be encountered by the wave function at some point. In the actual case, this happens around the $450$th period for incoming energies close to the sphaleron energy.}
\label{p}
\end{figure}

Although the Bloch-wave approach~\cite{Tye:2015tva} captures a key feature of the ($B+L$)-violating scattering process, two ingredients are not fully accounted for: \\
i) The quantum field theory problem reduces to a multidimensional quantum-mechanical(QM) problem, while the Bloch-wave approach focuses on the ($B+L$)-violating direction only. We have to include the ($B+L$)-conserving directions. Energy lost to the ($B+L$)-conserving directions is treated as some form of decay here.\\
ii) The quark and lepton masses effectively raise the energy of the final ground state with respect to the initial ground state (see Fig.~\ref{p}). This leads to a correction to the Bloch-wave solution that has to be taken into account.

Including these two effects, we find in this paper that $F(E)$ reaches a minimum at $E \simeq E_\text{sph}$, as shown in Fig.~\ref{FE}, and tends to stay there or even grow as $E$ continues to increase. Although we cannot determine the minimum value of $F(E)$, it is probably very small in our analysis, $F(E) \gtrsim 0$. The ``not-so-suppressed" exponential factor due to a very small $F(E/E_\text{sph} \simeq 1)$ may be ignored or be hiddenin our Bloch-wave analysis. This should be compared to the earlier estimate given in Ref.~\cite{Tye:2015tva} before the inclusion of the above two corrections, which estimated that $F(E/E_{0}) \to 0$ when $E$ reaches $E_\text{sph}$, as shown in Fig.~\ref{FE}. 

We know that $n$ can be identified with the Chern-Pontryagin index~\cite{tHooft:1976rip}. It is straightforward to treat the Chern-Simons variable $\mu(t)/\pi$ as a function of Minkowski time~\cite{Tye:2016pxi}. Although the choice of $\mu(t)$ is gauge dependent, it is gauge invariant at $\mu/\pi  \in {\bf Z}/2$ and can be identified with the topological Chern-Simons or Hopf index at $n=\mu/\pi  \in {\bf Z}$. 
Reference~\cite{Tye:2015tva} chose the coordinate $x={\mu}/{m_W}$ to mimic a spatial coordinate to obtain the one-dimensional time-independent Schr\"{o}dinger equation:
\begin{equation}
\label{Bloch}
  \left( - {1 \over 2m}{\partial ^2 \over \partial x^2} +V_0(x) \right) \Psi(x) = E\Psi(x).
\end{equation} 
where the potential $V_0(x)$ \cite{Manton:1983nd} and the mass $m$ \cite{Tye:2015tva} are, in the absence of the $U(1)$,
\begin{align}
\label{mainresult}
  V_0(x) &\simeq 4.75 \mbox{ TeV} \left(1.31\sin^2 (m_Wx)+0.60 \sin^4 (m_Wx) \right), \nonumber\\
  E_{sph} &= \mbox{max}  [V(x)] =V \left(\frac{\pi}{2 m_W} \right)= 9.11 \mbox{ TeV}, \nonumber \\
   m=&17.1 \mbox{ TeV},  
\end{align}
where the potential $V_0(\mu)\equiv V_0(x)$ is periodic, and we note that a rescaling of $x$ rescales $m$ without changing the physics. Since $\mu$ is gauge dependent when $2\mu/\pi \notin {\bf Z}$, the choice of extending $2\mu/\pi$ to noninteger values is a matter of convenience. [The choice $\mu/\pi - \sin (2 \mu)/2\pi$ is often used in the literature.] Choosing a different variable to interpolate between half-integers of $\mu/\pi$ will require a corresponding modification of $V_0(\mu)$ between the extrema as well as a modification of the mass $m$, which can become $x$ dependent~\cite{Tye:2016pxi}.

For the above periodic sphaleron potential $V_0(x)$, there is {\it a priori} no limit to $\Delta x = \pi \Delta n/m_W$, since the Bloch-wave passing bands run over all values of $x$.  Naively, this suggests that a single scattering seems capable of producing a large $\Delta n$, even for $\Delta n \to \infty$. However, the presence of fermions (with their zero modes) changes the picture in a fundamental way~\cite{tHooft:1976rip,tHooft:1976snw}. Here we first consider $\Delta n \ge 0$. In our analysis, the ground state with $n_B$ baryons is different from a ground state with  $n_B+ 3\Delta n$ baryons, so the sphaleron potential is no longer periodic, but rather is tilted upwards as $\Delta n$ increases. Reference~\cite{Tye:2017hfv} estimated that the potential $V(x)$ in Eq.~\eqref{Bloch} should take this into account, i.e., 
\begin{equation}
\label{cV}
 V_0(x) \to V(x)=V_0(x)+c\;x
\end{equation}
for $x> 0$, where $c \simeq 20 m_W/\pi$ GeV in the absence of Cabibbo-Kobayashi-Maskawa (CKM) mixing and $c\simeq 3 m_W/\pi$ GeV in the presence of CKM mixing, as illustrated in Fig.~\ref{p}. In this case, an incoming QM right-moving wave will be totally reflected, even if the wave function can penetrate multiple sphaleron barriers, so it seems that the ($B+L$)-violating process will not take place, since nothing stays at nonzero $\Delta n$. As a result, instead of $\Delta n \to \infty$, we seem to end up with $\Delta n=0$. 

In the actual situation, the quantum field theory problem translates to a multidimensional QM problem, including both ($B+L$)-violating and ($B+L$)-conserving directions. Energies diverted from the ($B+L$)-violating direction to the ($B+L$)-conserving directions will be treated as decay in the ($B+L$)-violating direction. We argue in this paper that, after taking into account the effect of decay and fermion masses, $\Delta n$ should be of the order of a few. A more accurate estimate probably requires a detailed study of the gauge dynamics. In summary, for energies close to the sphaleron energy, the incoming wave can decay after penetrating a few sphaleron barriers, so only part of the wave is reflected, and we expect that the ($B+L$)-violating processes will take place. If we start from the state with baryon number $B$, and reach the $B+ 3\Delta n$ state for $\Delta n >0$, the decay simply means that some energy goes to the baryon-number-conserving directions. This may be crudely approximated by the transition from one Bloch wave to a lower Bloch wave. For $\Delta n <0$, the wave ends up in a region with $B$ baryons and $3|\Delta n|$ antibaryons. The annihilation of $3|\Delta n|$ baryon-antibaryon pairs provides another decay channel. Energetically, a process like 
\begin{equation}\label{qqBL-1}
u_L + u_L \rightarrow e^+\mu^+ \tau^+ {\bar b} {\bar b} {\bar b}{\bar c}{\bar c} {\bar c} {\bar u} + X
\end{equation}
may be more likely than the above $\Delta n =+1$ process in Eq.~\eqref{qqBL+1}.

\section{Review}
\label{BlochR}
  
Bloch waves are solutions to Eqs.~\eqref{Bloch} and~\eqref{mainresult} (see below for the reason why continuous Bloch-wave bands exist). There are 148 such conducting bands below the sphaleron energy $E_\text{sph}=9.11$ TeV~\cite{Tye:2015tva}. The lowest one is at $0.3421$ TeV with an exponentially small width ($\Gamma \sim 10^{-180}$ TeV), while the one just below $E_\text{sph}$ is at $9.081$ TeV with width $\Gamma \simeq 7.2$ GeV. The one just above $E_\text{sph}$ is at $9.113$ TeV with width $\Gamma \simeq 15.6$ GeV. This is evaluated in the absence of fermions. 
 
In the presence of left-handed fermions, the periodic potential is no longer exactly periodic and the Bloch-wave direction we are interested in is different than the usual $|\theta\rangle$ vacuum direction. At the classical level, there exist $n_L=12$ ($i=1,2,...,n_L$) globally conserved $U(1)$ currents for the (left-handed) quark and lepton electroweak doublets,
\begin{equation}
\label{JLi}
J^{(i) \mu}={\bar \Psi}^{(i)}_L \gamma^{\mu}\Psi^{(i)}_L
\end{equation} 
corresponding to the conservation of fermion number. However, this conservation is broken by the presence of an anomaly~\cite{Adler:1969,Bell:1969ts},
\begin{equation}
 \label{JLK}
\partial_{\mu} J^{(i) \mu} = \frac{g^2}{16 \pi^2} \Tr\left[F_{\mu \nu} \tilde{F}^{\mu \nu}\right] = \partial_{\mu} K^{\mu}
\end{equation} 
where $\tilde{F}^{\mu \nu}$ is the dual of  ${F}^{\mu \nu}$ and there exists a (gauge-dependent) current $K^{\mu}$. 
In the presence of instanton solutions in Euclidean spacetime~\cite{Belavin:1975fg}, 
\begin{equation}
\label{chern} N =\frac{g^2}{16 \pi^2} \int d^4x \Tr\left[F_{\mu \nu} \tilde{F}^{\mu \nu} \right], 
\end{equation} 
where the topological index $N$ takes only integer values. An instanton with value $N$ leads to the tunneling process $\left|n \right\rangle \rightarrow \left|n+N \right\rangle$.

\begin{figure}
	\centering
	\includegraphics[width=6cm]{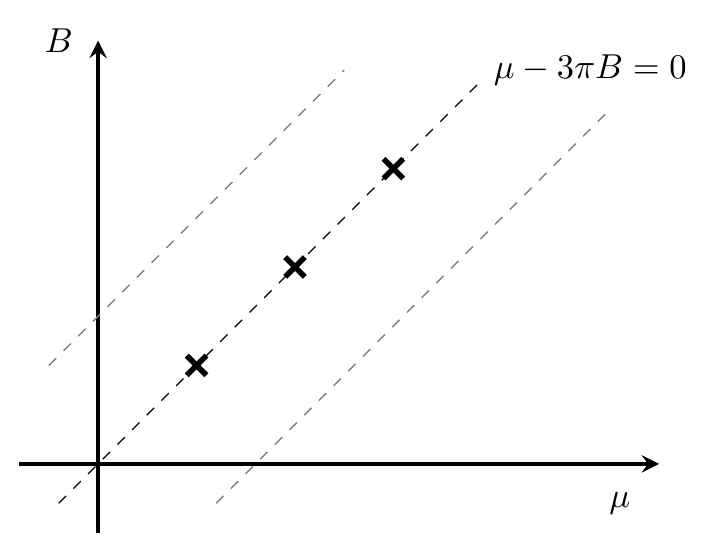}
	\caption{States in the $(\mu-3\pi B)$-conserving direction. Tunneling in the EW theory is along the diagonal direction while the $|\mu\rangle$s in the QCD $|\theta\rangle$ are along the horizontal direction. }
	\label{muBplot}
\end{figure}

One can construct a (gauge-dependent) conserved current ${\bar J}^{i, \mu}$ and the corresponding conserved charge ${\cal Q}^{i}$,
\begin{eqnarray}
%&J_F^{\mu}=\frac{1}{n_L} \sum_{i=1}^{n_L}J^{(i) \mu}  \nonumber\\
&\partial_{\mu} {\bar J}^{i,\mu} = \partial_{\mu}(K^{\mu}-J^{i,\mu}) =0,   \nonumber\\
& {\cal Q}^i = \int d^3 x {\bar J}^{i, 0} =  Q_G -Q_F^i
\label{conservQ}
\end{eqnarray} 
which is the winding number $Q_G$ of the gauge field minus the normalized $i$th fermion doublet number $Q_F^i$. 
So a state may be described by $n_L+1$ values,
$|n\rangle =|n_G, n_F^{(i)}\rangle$, with $Q_G|n \rangle = n_G|n \rangle$ and $Q_F^i |n \rangle = n_F^{(i)}|n \rangle$.
Let us start with a vacuum state $| 0\rangle = |0,0,...,0\rangle$; then, a tunneling transition preserving all ${\cal Q}^i$ means 
$$| 0\rangle \equiv | 0, 0\rangle \rightarrow | n_G, n_F\rangle = |N=B, 3B+3L\rangle$$
which has net baryon number (i.e., baryon minus antibaryon number) $B$ and net lepton number $L$, and $n_F= n_F^{(1)}=n_F^{(2)}=...=n_F^{(n_L)}$.  It is important to emphasize that the $|n_G, n_F \ne 0\rangle$ states 
are obviously not vacua, but these classical ground states are almost degenerate with the vacuum states for very soft massless fermions and for not too big $n_F$. More generally, a ($B+L$)-violating process with $\Delta n \ne 0$ refers to such a ${\cal Q}$-conserving transition: 
\[|n_G, n_F^{(i)} \rangle \rightarrow |n_G+\Delta n, n_F^{(i)}+\Delta n \rangle \]
The $\mu$ direction in the Schr\"odinger equation of Eq.~\eqref{Bloch} refers to this ${\cal Q}^{i}$-conserving direction, as shown in Fig.~\ref{muBplot}. 

\begin{figure}
\includegraphics[width=6.5cm]{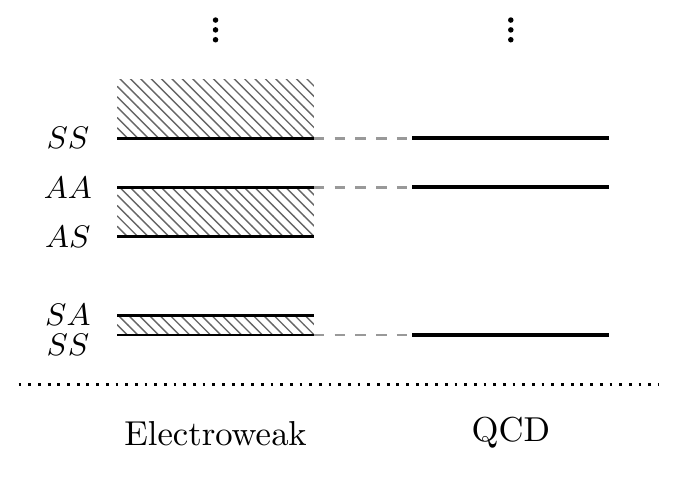}
\caption{Comparison between EW theory and QCD \cite{Bachas:2016ffl,Tye:2017hfv}. The edges of the Bloch pass bands are labeled by $(S(A)S(A))$, where the first letter denotes symmetric ($S$, even) or antisymmetric ($A$, odd) with respect to the bottom of potential and the second denotes that with respect to the top of the potential~\cite{Tye:2015tva}. EW theory has Bloch wave bands but no $|\theta \rangle$ vacuum, while QCD has the $|\theta \rangle$ vacuum but no Bloch-wave bands.}
\label{fig:bands}
\end{figure}

A Bloch-wave state takes the form (for integer $\pi \mu \in {\bf Z}$)
\begin{equation}
\label{kEW}
|k \rangle= \sum_{\mu} e^{ik\mu/\pi}|n_G +\mu/\pi, n_F^{(i)}+\mu/\pi \rangle 
\end{equation}
where $\mu=xm_W$ is the spatial QM coordinate in the Schr\"odinger equation, which has  continuous Bloch-wave bands. (See Appendix A for further explanation.) Clearly, $|k \rangle$ is very different from the standard QCD $\theta$ vacuum,
\begin{equation}
\label{thetaQCD}
|\theta \rangle =  \sum_{\mu} e^{i\theta \mu/\pi}|n_G +\mu/\pi, n_F^{(i)}=0 \rangle
\end{equation}
Note that QCD has a $\theta$ vacuum but no Bloch-wave bands, while EW theory has Bloch-wave bands but no $\theta$ vacuum~\cite{Bachas:2016ffl,Tye:2017hfv}. We can see this clearly in the one-dimensional QM setup. For a periodic potential, the wave function is given by $\Psi(x)= e^{ikx} u_k(x)$, where $u_k(x+\pi/m_W)=u_k(x)$ is periodic.  Bloch-wave band solutions appear for ranges of $k$, as illustrated in Fig.~\ref{fig:bands}. For QCD, each $|n\rangle$ is a gauge-transformed version of $|n=0\rangle$, so they are physically identical; that is $\Psi(x)$  itself must be periodic, or $k=2m_W$, so only $k=2nm_W$ in a  Bloch-wave band is allowed. This happens for $SS$ and $AA$ edges of the Bloch-wave bands, as shown in Fig.~\ref{fig:bands}. Pictorially, the QCD periodic potential is like that of a rigid pendulum in the presence of an external (e.g., gravitational) field, where $2xm_W$ measures the angle and a rotation of $2\pi$ implies $\Psi(x)=\Psi(x+\pi/m_W)$.  Here the energy levels have no width, so a transition between them is absent in this approximation.

If instead of a periodic potential one considers a potential with $N$ identical barriers, then each continuous band is replaced by a set of  $N-1$ discrete energies, still separated by gaps. For $N=2$, each band is reduced to a single energy level, revealing the resonant tunneling phenomenon.  For a single ($N=1$) barrier, the tunneling suppression on the rate goes like $e^{-4\pi/\alpha}$. This is exponentially suppressed even in QCD, where $\alpha_\text{QCD}(m_Z) \simeq 0.12$; so the tunneling time scale would be orders of magnitude longer than a typical QCD scattering process. It is the resonant tunneling effect that enables the tunneling through barriers unsuppressed, which leads to the $|\theta \rangle$ vacuum required by cluster decomposition.
	
Also note that there are ($B+L$)-conserving directions $y_i$ where no barrier penetration is involved. However, as is clear from Eqs.~\eqref{kEW} and ~\eqref{thetaQCD}, they are orthogonal to the barrier penetration direction $x$ and do not appear in the construction of either $|k\rangle$ or $|\theta\rangle$.

\section{Resonant Tunneling with Decay}
\label{ResD}

In the one-dimensional QM problem setup in Eqs.~\eqref{Bloch},~\eqref{mainresult}, and~\eqref{cV}, we know that an incoming wave from the left will penetrate a number of sphaleron barriers, but then will hit a barrier higher than the incoming energy and be reflected back, as shown in Fig.~\ref{p}. The reflection coefficient turns out to be unity, so this seems to imply that no ($B+L$)-violating process happens, as alluded to in Ref.~\cite{Bachas:2016ffl}. This may be true for the lowest Bloch wave function, but not for the higher Bloch-wave functions that we are interested in, especially ones that are close to or above the sphaleron height, which can decay [by losing energy to the ($B+L$)-conserving directions] to the lower Bloch-wave functions. This decay plays a similar role as tunneling through another barrier, which would allow the presence of resonance enhancement. It is amazing that this resonant tunneling plus decay phenomenon was employed long ago in the famous triple-$\alpha$ transition to form carbon in nucleosynthesis in stars in the early Universe.

A particle coming from the left with energy $E$ below the barrier heights will tunnel through the barriers. Here the incoming energy $E$ is energy along the ($B+L$)-violating direction. In the WKB approximation, the connection matrix for amplitudes on the two sides of the $i$th potential barrier is
\beq
M_i=
\begin{pmatrix}
	\cosh{\hat{S}_i} & i\sinh{\hat{S}_i}\\[0.7em]
	-i\sinh{\hat{S}_i} & \cosh{\hat{S}_i}
\end{pmatrix}\;,
\eeq
and particles propagate over the subsequent classically allowed region with connection matrix
$\Phi_i$,
\beq
\Phi_i=
\begin{pmatrix}
	e^{-iL_i} & 0\\
	0 & e^{iL_i}
\end{pmatrix}\;,
\eeq
where
\beq
\hat{S}_i = \int_{a_i}^{b_i} \sqrt{2m(V(x) -E)} \rmd x + \ln 2\;,
\eeq
and 
\beq
L_i = \int_{b_i}^{a_{i+1}} \sqrt{2m(E-V(x))} \rmd x\;.
\eeq
$a_i$ and $b_i$ are the turning points. The general formula for going through $m$ barriers is 
\beq
M_{\text{total}}=\left(\prod_{i=1}^{m-1}M_i\Phi_i\right) M_m\;.
\eeq
In the absence of decay, unitarity requires that the determinant $|\det M_i|=1$.

For the periodic sphaleron potential $V_0(x)$, where all $M_i$ and $\Phi_i$ are identical and $m \to \infty$, the solution is the Bloch waves, which are passing bands with continuously allowed energies within each band and adjacent passing bands are separated by (disallowed) gaps. For finite $m$, each pass band contains a discrete set of allowed energies, which becomes a ``dense discretuum" for large but finite $m$. Here, to get an idea of the likely value of $|\Delta n|$, we shall implement the decay as well as the tilting of the potential $V(x)$ in the cases with one, two and three barriers. 

To describe a decaying state, usually we can add a negative imaginary part to the eigenenergy, which gives $\Psi(x,t)=\phi(x)\exp{\left[-i(E-i\gamma/2)t/\hbar\right]}$. Because we are calculating amplitudes of initial states from connection matrix multiplying final states, this is actually the time-reversal process. Since time-reversal operator is antiunitary, $\hat{T}\Psi(x,t)=\phi^*(x)\exp{\left[-i(E+i\gamma/2)t/\hbar\right]}$, the probability is $|\hat{T}\Psi(x,t)|^2=|\phi(x)|^2\exp{(\gamma t/\hbar)}$, as expected for decaying behavior in the time-reversal process. As a result, we describe decaying behavior here by letting the eigenenergy acquire a positive imaginary part,
\beq
L_i \;\to\; \frac{1}{\hbar}\int_{b_i}^{a_{i+1}}\sqrt{2m\left(E-V(x)+i\frac{\gamma}{2}\right)} \rmd x\;.
\eeq
This is approximately $L=l+i\Delta$, where $l$ represents the real part of the integral. Usually, $\gamma \ll E$, so we can obtain an approximate expression for $\Delta$,
\beq
\Delta_i\approx\frac{\gamma}{4\hbar}\int_{{b_i}'}^{{a_{i+1}}'}\sqrt{\frac{2m}{E-V(x)}}\rmd x\;,
\label{eq:delta}
\eeq
where ${b_i}'$ and ${a_{i+1}}'$ are new limits of the integral of the imaginary part, which are determined by
\beq
E-V({b_i}')=E-V({a_{i+1}}')=\frac{\gamma}{2}\;.
\eeq
The reason that we use these limits for the integral of the imaginary part is that WKB method fails near the turning point, where the contribution to $\Delta_i$ is very small compared to that from the middle region, where $E-V(x)\gg\frac{\gamma}{2}$. With the help of a general formula, we can discuss some specific cases.

\begin{figure}
	\centering
	\includegraphics[width=7.8cm]{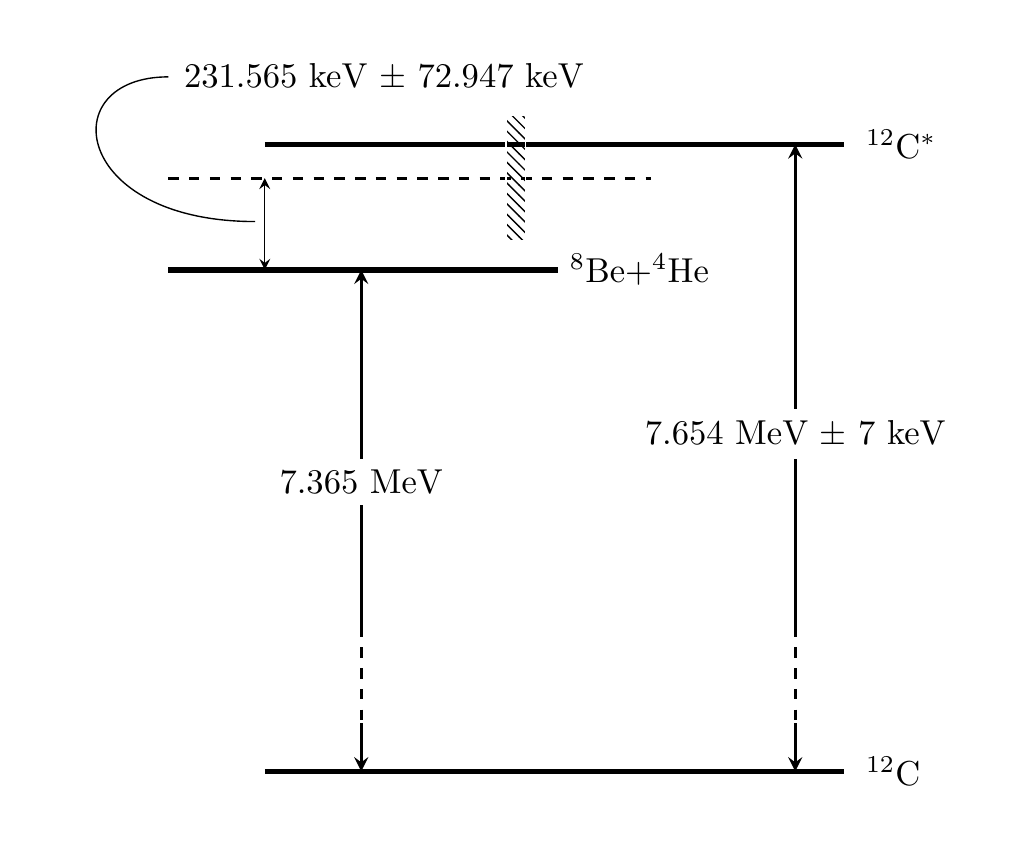}
	\caption{Energy levels of different nuclei. The dashed line is the position of the fusion window for a nuclear reaction between $^4$He and $^8$Be. The shaded area represents the width of the fusion window, which includes the Hoyle state.}
	\label{fig.energylevels}
\end{figure}

\section{Single-Barrier Case}
\label{single}

It is amazing that this simplest case (one tunneling channel and one decay channel) was actually applied to solve a major puzzle in nucleosynthesis. Before the 1950s, cosmologists and astrophysicists could not find a way to produce carbon nucleus and beyond, either primordially or in stars. Then, Salpeter proposed~\cite{Salpeter1952} and Holye reinforced~\cite{Hoyle1953} the idea that the carbon nucleus could be formed if it  has an excited state at a particular energy, as shown in Fig.~\ref{fig.energylevels}, which was subsequently discovered at precisely the energy predicted. (A brief review is included in Appendix B). Let us go over this case as a step towards explaining the ($B+L$)-violating process.

\begin{figure}
	\centering
	\includegraphics[height=5cm]{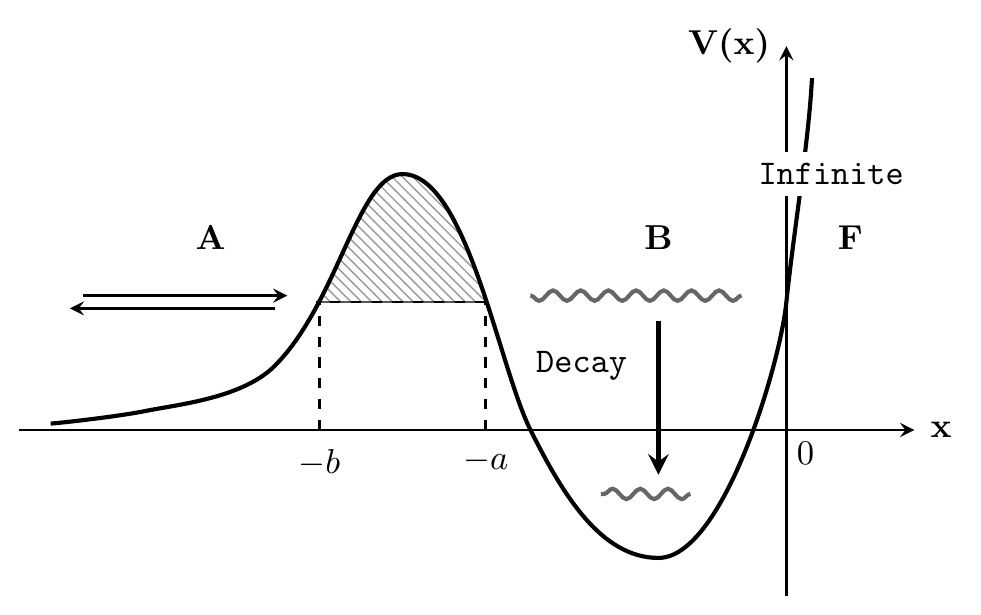}
	\caption{Shaded region is inaccessible classically.}
	\label{pic1}
\end{figure}

As shown in Fig.~\ref{pic1}, we consider that the second barrier is simply an infinite wall. In the absence of decay, the reflection amplitude is given by
\beq
\mathcal{R} = \frac{t+e^{2iL}}{1+t e^{2iL}} \to |{\mathcal R}|^2=1, \quad t\equiv\tanh{\hat{S}}\;,
\eeq
which means that the particle is totally reflected. Nothing ends in region $\bf B$. Suppose that the particle can decay to a lower energy level in region $\bf B$, say the ground state. In the triple-$\alpha$ case, carbon-12 decays via the transition from the excited carbon-12 to a lower (or the ground) state and emits photons, while the decay in the sphaleron case is the transition from a higher band to a lower band via a loss of energy to the  ($B+L$)-conserving direction.

According to the discussion in the last section, we let $L=l+i\Delta$ and define $\beta\equiv e^{-2\Delta}$ as the decaying parameter in region \textbf{B}, which gives
\beq
|\tilde{\mathcal{R}}|^2=\frac{(\frac{\beta}{t}+
	\frac{t}{\beta})+2\cos(2l)}{(t\beta+\frac{1}{t\beta})+2\cos(2l)}\;.
\eeq
The probability of staying in region $\bf B$ is now given by
\beq
|\mathcal{G}|^2\equiv1-|\tilde{\mathcal{R}}|^2=\frac{(\frac{1}{\beta}-\beta)(\frac{1}{t}-t)}{\frac{1}{t\beta}+t\beta+2\cos(2l)}\;.
\label{coefficient}
\eeq
\begin{figure}
\centering
\includegraphics[width=7.8cm]{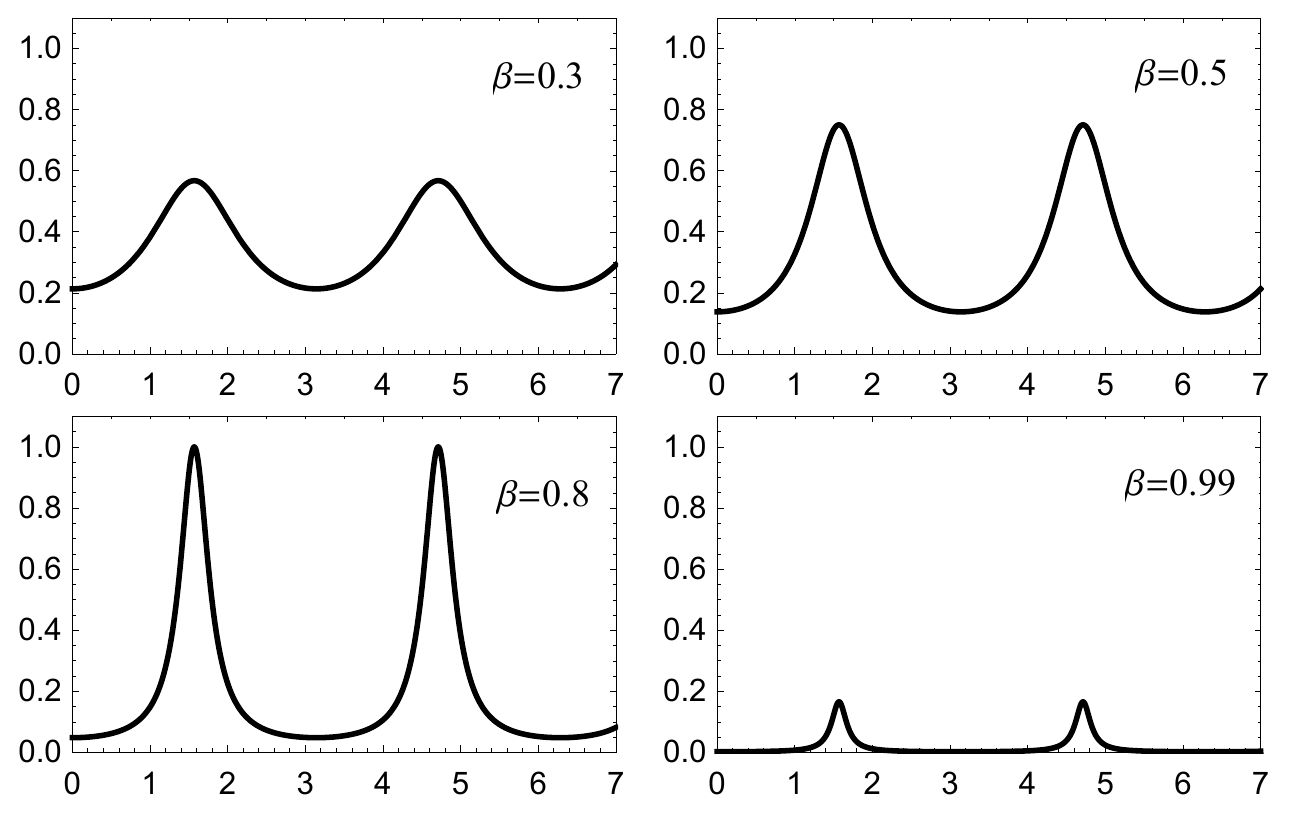}
\caption{The probability $|\mathcal{G}|^2$ of ending up in region {\bf B} in Fig.~\ref{pic1} as a function of the incoming energy $E$ at different values of the parameter $\beta$ with a fixed $t=0.8$. A peak appears as $E$ hits one of the resonance energies besides the ground state.}
\label{fig.singleD}
\end{figure}
If $\beta=1$, which means there is no decay, $|\mathcal{G}|^2=0$ since the reflection probability $|{\mathcal R}(\beta=1)|^2=1$, irrespective of the incoming energy.  For $\beta<1$, the probability of staying in region $\bf B$ shows a resonance pattern for $|\mathcal{G}|^2$ as a function of the incoming energy, as plotted in Fig.~\ref{fig.singleD}. For certain choices of the parameters $(t,\beta)$, the probability $|\mathcal{G}|^2$ can reach 1 at resonance energies. We focus on a specific resonance energy level $E_1$. we expand the $|\mathcal{G}|^2$ around this level,
\beq
|\mathcal{G}|^2 \simeq \frac{(\beta^2-1)(t^2-1)}{(t\beta-1)^2+
	4t\beta[\left.(\frac{\partial l}{\partial E})\right|_{E=E_1}]^2(E-E_1)^2}.
\eeq
Furthermore, we can let 
\begin{align}
	\Gamma_a&=\frac{1}{2\sqrt{t\beta\omega}}
	\left(1-t\beta+|t-\beta|\right)\nonumber\\
	\Gamma_b&=\frac{1}{2\sqrt{t\beta\omega}}
	\left(1-t\beta-|t-\beta|\right)\nonumber\\
	\Gamma&=\Gamma_a+\Gamma_b,\nonumber
\end{align}
where $\omega=\left[ \left. \left(\frac{\partial l}{\partial E}\right)\right|_{E=E_1}\right]^2$. The probability $|\mathcal{G}|^2$ is now a function of the incoming energy
\beq
\label{eq.decay}
|\mathcal{G}(E)|^2 =\frac{\Gamma_a\Gamma_b}{(\Gamma/2)^2+(E-E_1)^2}.
\eeq
which is the Breit-Wigner formula. Here $\Gamma$ is the total width of a certain level and $\Gamma_{a,b}$ are the two different channels: one decays to the ground state in region $\bf B$ and the other decays back to the initial state.

The total reflection amplitude consists of two parts in this case. One is a direct reflection after hitting the barrier from the left with amplitude $\mathfrak{r}$, and the other is a tunneling back out to the left (region $\bf A$) from region $\bf B$, with amplitude $\mathcal{D}_{-}$. The unitary relation is
\beq
|\tilde{\mathcal{R}}|^2+|\mathcal{G}|^2=|\mathfrak{r}+\mathcal{D}_{-}|^2+|\mathcal{D}_{g}|^2=1.
\eeq
The two channels of resonance decay in region {\bf B} could be interpreted as one via tunneling with amplitude $\mathcal{D}_{-}$, and the other with decay to lower energy levels with probability  $|\mathcal{D}_{g}|^2$.

\section{Double-Barrier Case}
\label{Double}

\begin{figure}
	\centering
	\includegraphics[height=3.8cm]{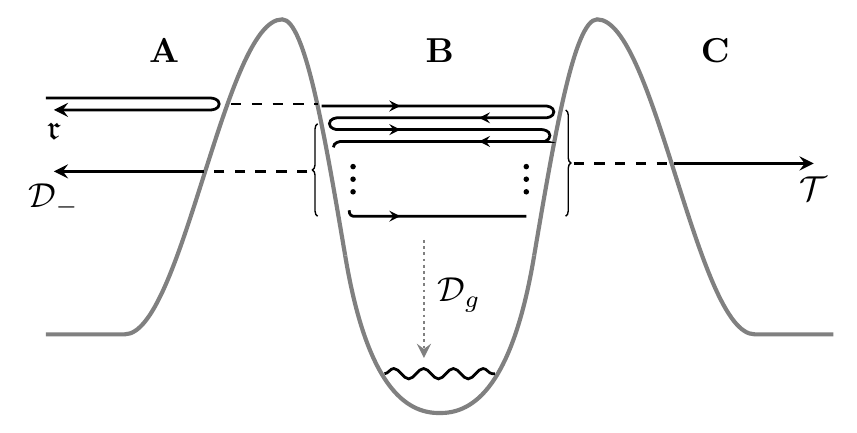}
	\caption{Resonant tunneling through a double-barrier potential.}
	\label{barriers}
\end{figure}

Next, we would like to consider the two-barrier case, as shown in Fig.~\ref{barriers}. As usual, the direct tunneling amplitude $\mathcal{T}$ from the left (region $\bf A$) to the right (region $\bf C$) is doubly exponentially suppressed unless the energy is close to a resonance in region $\bf B$. 
In this case, a resonant state in region $\bf B$ can decay in three ways:\\
i) Decay to a lower energy state in region $\bf B$ with amplitude $\mathcal{D}_{g}$.\\
ii) Tunnel back with amplitude $\mathcal{D}_{-}$, contributing to the reflection amplitude $\mathcal{R}$.\\
iii) Tunnel forward to the right with amplitude $\mathcal{T}$. 

With an incoming wave from the left, unitarity demands
\beq
|\mathcal{R}|^2+|\mathcal{G}|^2+|\mathcal{T}|^2=|\mathfrak{r}+\mathcal{D}_{-}|^2+|\mathcal{D}_{g}|^2+|\mathcal{T}|^2=1\;,
\eeq
with the assumption that there exists a number of decay channels of the particular resonance $|\mathcal{G}|^2 = \sum_j |\mathcal{D}_{gj}|^2$. By hitting the resonance, $|\mathcal{T}|^2$ and $|\mathcal{G}|^2$ can be unsuppressed.

If we consider a wave packet going through a double-barrier potential with a certain energy near the resonance in region $\bf B$, the transmission from region \textbf{A} to region \textbf{C} can be interpreted as a decay to another state, which can precisely reproduce the exponential decay law of an unstable state. The total transmission probability is $\gamma/{\Delta E}$, where $\gamma$ stands for the decay width of the state in region {\bf A} and $\Delta E$ is the energy spread of this wave packet. Without the decay $\mathcal{D}_g$, this interpretation mimics the single-barrier case in the last section. This tells us that ``multi-escaping-channels" for middle unstable state will introduce the resonance enhancement effect.

\begin{figure}
	\centering
	\includegraphics[width=6.5cm]{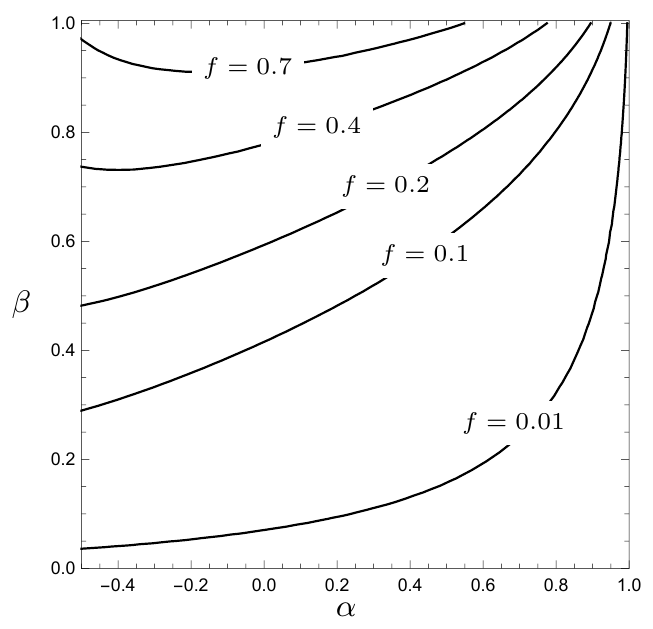}
	\caption{The function $f(\alpha,\beta)\equiv\max{|\mathcal{T}|^2}$ in the $(\alpha, \beta)$ parameter space. Maximum tunneling $\max{|\mathcal{T}|^2}=1$ for $\alpha=0$ (same barrier) and $\beta=1$ (no decay).}
	\label{fig:maxT}
\end{figure}

The connection formula in this case is
\beq
M=
\begin{pmatrix}
	c_1 c_2 e^{-iL} + s_1 s_2 e^{iL} & i \big(c_1 s_2 e^{-iL}+s_1 c_2 e^{iL}\big)\\[0,6em]
	-i \big(s_1 c_2 e^{-iL}+ c_1 s_2 e^{iL}\big) & s_1 s_2 e^{-iL}+ c_1 c_2 e^{iL}
\end{pmatrix}\;,
\eeq
where $s_i=\sinh{\hat{S}_i}$ and $c_i=\cosh{\hat{S}_i}$. To include decaying behavior, we let $L=l+i\Delta$. For simplicity, we define $t_i=\tanh{\hat{S}_i}$ and $\beta=e^{-2\Delta}$. The transmission probability can be calculated directly,
\beq
|\mathcal{T}|^2=\frac{\beta (1-t_1^2)(1-t_2^2)}{1+(t_1t_2 \beta)^2+2t_1t_2\beta \cos{2l}}\;.
\eeq
For $\beta=1$, $t_1=t_2$, and $\cos{2l}=-1$, the transmission probability $|\mathcal{T}|^2=1$. The condition for resonance enhancement is $l=(n+\frac{1}{2})\pi$, where $n\in\mathbf{N}$. At the appropriate incoming energy, the transmission can reach a maximum that is not doubly exponentially suppressed. An interesting situation is when the second barrier is slightly higher than the first one, $\hat{S}_2>\hat{S}_1$. Is we let $\hat{S}_2=\hat{S}_1+\Delta{S}$, we have the relation
\beq
t_2=\tanh{\hat{S}_2}=\frac{\tanh\hat{S}_1+\tanh\Delta{S}}{1+\tanh{\hat{S}_1}\tanh{\Delta{S}}}=\frac{t_1+\alpha}{1+t_1 \alpha}\;,
\eeq
where $\alpha=\tanh{\Delta{S}}<1$. We use this relation to replace $t_2$ in the transmission probability and define the function
\beq
f(\alpha,\beta)\equiv\max{|\mathcal{T}|^2}=\frac{\beta(1-{t_1}^2)^2(1-\alpha^2)}{(1-\beta{t_1}^2+(1-\beta)t_1 \alpha)^2}\;.
\eeq
As shown in Fig.~\ref{fig:maxT}, when $\alpha=0$ the two barriers are the same and the maximum of $|\mathcal{T}|^2$ could reach unity if $\beta=1$. As $\alpha$ increases, the height of second barrier grows and $\max{|\mathcal{T}|^2}$ decreases. If $\alpha\to1$, which means that the second barrier becomes impossible to penetrate, the transmission probability vanishes, as expected. The decay probability $|\mathcal{G}|^2$ shares the same features as $|\mathcal{T}|^2$. The potential tilt reduces the $\max{|\mathcal{T}|^2}$, while the resonance enhancement pattern preserves.

\section{Triple-Barrier Case}
\label{Triple}
	
In this section we extend the above analysis to the three-barrier case and use seven parameters describe the situation, which are $t_{1,2,3}$ for three-barrier tunneling, $\beta_{1,2}$ for decay between barriers, and $l_{1,2}$ for the energy levels in the two potential wells. The explicit expressions are given as
\beq
|\mathcal{T}|^2=\frac{1}{D}\left(\beta_1\beta_2\left(1-{t_1}^2\right)\left(1-{t_2}^2\right)\left(1-{t_3}^2\right)\right)\;,
\nonumber
\eeq
\vspace{-0.4cm}
\begin{align*}
|\mathcal{R}|^2=&\frac{1}{D}\bigg({t_1}^2+\left(t_2\beta_1\right)^2+\left(t_1t_2t_3\beta_2\right)^2+\left(t_3\beta_1\beta_2\right)^2\\
&+2t_1t_2\beta_1\left(1+{t_3}^2{\beta_2}^2\right)\cos{(2l_1)}\\
&+2t_2t_3\beta_2\left({t_1}^2+{\beta_1}^2\right)\cos{(2l_2)}\\
&+2t_1{t_2}^2t_3\beta_1\beta_2\cos{[2(l_1-l_2)]}\\
&+2t_1t_3\beta_1\beta_2\cos{[2(l_1+l_2)]}\bigg)\;,
\end{align*}
\vspace{-0.4cm}
\begin{align*}
|\mathcal{G}|^2=&\frac{1}{D}\bigg(1+\left(t_1t_2\beta_1\right)^2+\left(t_2t_3\beta_2\right)^2+\left(t_1t_3\beta_1\beta_2\right)^2\\
&-\beta_1\beta_2\left(1-{t_1}^2\right)\left(1-{t_2}^2\right)\left(1-{t_3}^2\right)\\	&-\left({t_1}^2+\left(t_2\beta_1\right)^2+\left(t_1t_2t_3\beta_2\right)^2+\left(t_3\beta_1\beta_2\right)^2\right)\\	&+2t_2t_3\beta_2\left(1-{t_1}^2\right)\left(1-{\beta_1}^2\right)\cos{(2l_2)}\bigg)\;,
\end{align*}
\vspace{-0.4cm}
\begin{align*}
D=&1+\left(t_1t_2\beta_1\right)^2+\left(t_2t_3\beta_2\right)^2+\left(t_1t_3\beta_1\beta_2\right)^2\\
&+2t_1t_2\beta_1\left(1+{t_3}^2{\beta_2}^2\right)\cos{(2l_1)}\\
&+2t_2t_3\beta_2\left(1+{t_1}^2{\beta_1}^2\right)\cos{(2l_2)}\\
&+2t_1{t_2}^2t_3\beta_1\beta_2\cos{[2(l_1-l_2)]}\\
&+2t_1t_3\beta_1\beta_2\cos{[2(l_1+l_2)]}\;,
\end{align*}
where $D$ is the denominator. In the simple case where $t=t_1=t_2=t_3$, $\beta_i=1$ and $l=l_1=l_2$, 
the transmission probability becomes
$$|\mathcal{T}|^2=\frac{(1-{t}^2)^3}{1+5t^2+4t^2(1+t^2)\cos(2l) +2t^2 \left(2 (\cos(2l))^2-1\right)}$$
which yields $|\mathcal{T}|^2=1$ for $\cos(2l) =  - (1+t^2)/2$. The analytical expressions for the probabilities are too complicated, and numerical calculations would be more helpful. Here the calculation could not distinguish the probabilities of the decays into the two potential wells. Thus, $|\mathcal{G}|^2$ refers to the total decay probability that is neither reflected nor transmitted.

\section{Sphaleron Potential}
\label{multiple}

If there are two identical barriers only and no decay, when the incoming energy hits the resonant energy, the transmission probability approaches unity, due to the coherent sum of an infinite set of paths, even if the tunneling through a single barrier is exponentially suppressed. For energies away from the resonant energy, the transmission probability is typically doubly exponentially suppressed and the reflection dominates. The process could be understood as tunneling through the first barrier while the second tunneling is treated as the decay of the resonant state. So, if we raise the height of the second barrier substantially to infinity in which the particle cannot reach region {\bf C}, the reflection probability approaches unity; no particle can stay in region $B$, even if now we lower the first barrier height so it becomes easier to reach region {\bf B} from region {\bf A}.

We apply this understanding to the ($B+L$)-violating process in the Bloch-wave approach. If we include the fermion masses, sooner or later the wave will be stopped by a totally inaccessible barrier as shown in Fig.~\ref{p}, causing the reflection probability to be unity without decay, which implies that no ($B+L$) violation takes place. As discussed in previous cases, tunneling plus decay may produce a similar enhancement as the resonant tunneling phenomenon. The idea clearly applies to the more complicated cases, so an estimate of the rate may be nontrivial. A typical case will involve multiple tunneling and decay channels, which can also produce a resonant enhancement effect. 	
\begin{figure}
	\centering
	\includegraphics[width=7cm]{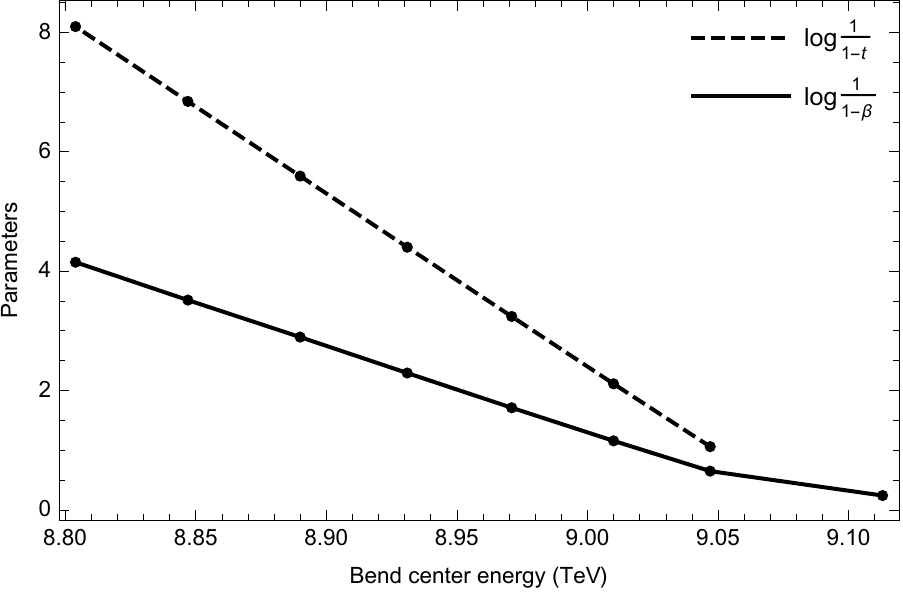}
	\caption{The two parameters $(t,\beta)$ describing the tunneling and decay behaviors tend to merge as Bloch band center energy increases.}
	\label{bandparameter}
\end{figure}

Due to the lack of exact information about the decay behaviors of the resonance level in the sphaleron case, we can only make order-of-magnitude estimations. In the pure Bloch-wave-model with only massless fermions, we have the total transmission probability for a wave packet described as 
\[
	|\mathcal{T}|^2=\frac{\Gamma_{B}}{\Gamma_{B}+0.5(\Delta_{+}+\Delta_{-})}\;,
\] 
where $\Gamma_{B}$ is the Bloch-wave band width and $\Delta_{\pm}$ represents the band gap above or below the band. The denominator can be considered as the energy spread of a wave packet $\Delta E$. Physically, one expects that a high energy state would have a large decay width, which is also the case for the Bloch-band width. It is a reasonable approximation to use $\Gamma_B$ (calculated in Ref.~\cite{Tye:2015tva}) as input $\gamma$ when calculating the decay parameters $\beta_i$ with Eq.~\eqref{eq:delta}. We use the modified potential in Eq.~\eqref{cV} (in which an additional linear term is included due to the accumulated fermion mass) to calculate the tunneling parameter $t_i$ for each barrier. With the decay behavior considered, the possibility that an incoming wave tunnels to other states could be nonzero.

If we only consider one decay channal without transmission (the second barrier is infinitely high), we have two parameters ($t,\beta$) to describe the system. We can see from Eq.~\eqref{eq.decay} that the enhancement effect would be more explicit if these two parameters are close to each other. As shown in Fig.~\ref{bandparameter}, resonance enhancement are expected as energy approaches sphaleron energy (9.1 TeV).
\begin{table}
\caption{\label{tab:rlevel}Resonance in a potential well.}
\begin{ruledtabular}
\begin{tabular}{lccc}
	\multicolumn{4}{c}{2 Identical Barriers}\\
	\hline
	\multicolumn{2}{l}{$E_0$ (TeV)} 	& 	\multicolumn{2}{c}{$\Gamma$(TeV)} \\ 
	\hline
	\multicolumn{2}{l}{9.080}	&\multicolumn{2}{c}{0.09793}\\ 
	\multicolumn{2}{l}{9.053}	&\multicolumn{2}{c}{0.02669}\\ 
	\multicolumn{2}{l}{9.019}	&\multicolumn{2}{c}{$3.031\times10^{-3}$}\\ 
	\multicolumn{2}{l}{8.983}	&\multicolumn{2}{c}{$2.790\times10^{-4}$}\\ 
	\multicolumn{2}{l}{8.945}	&\multicolumn{2}{c}{$2.228\times10^{-5}$}\\ 
	\multicolumn{2}{l}{8.905}	&\multicolumn{2}{c}{$1.592\times10^{-6}$}\\
	\multicolumn{2}{l}{8.864}	&\multicolumn{2}{c}{$1.036\times10^{-7}$}\\
	\multicolumn{2}{l}{8.821}	&\multicolumn{2}{c}{$6.206\times10^{-9}$}\\
	\multicolumn{2}{l}{8.779}	&\multicolumn{2}{c}{$3.454\times10^{-10}$}\\
	\multicolumn{2}{l}{$\vdots$}&\multicolumn{2}{c}{$\vdots$}\\
	\hline
	\hline
	\multicolumn{4}{c}{$c=3 m_W/\pi$ GeV}\\
	\hline
	\multicolumn{2}{c}{Resonance in $n=1$}& \multicolumn{2}{c}{Resonance in $n=2$}\\
	\hline
	$E_1$ (TeV) 	& 	$\Gamma_1$(TeV) 	& 	$E_2$(TeV)		& 	$\Gamma_2$(TeV) 	\\ \hline
	9.083	&	0.1016				&9.086	&0.1008		 \\ 
	9.056	&	0.02690				&9.059	&0.0268		 \\ 
	9.022	&$3.045\times10^{-3}$	&9.025	&$3.045\times10^{-3}$	\\ 
	8.986	&$2.804\times10^{-4}$	&8.989	&$2.804\times10^{-4}$	\\ 
	8.948	&$2.239\times10^{-5}$	&8.951	&$2.239\times10^{-5}$	\\ 
	8.908	&$1.600\times10^{-6}$	&8.911	&$1.599\times10^{-6}$	\\
	8.867	&$1.041\times10^{-7}$	&8.869	&$1.041\times10^{-7}$	\\
	8.825	&$6.237\times10^{-9}$	&8.828	&$6.237\times10^{-9}$	\\
	8.782	&$3.472\times10^{-10}$	&8.785	&$3.472\times10^{-10}$	\\
	$\vdots$&$\vdots$				&$\vdots$	&$\vdots$	\\
	\hline
	\hline
	\multicolumn{4}{c}{$c=20 m_W/\pi$ GeV}\\
	\hline
	\multicolumn{2}{c}{Resonance in $n=1$}& \multicolumn{2}{c}{Resonance in $n=2$}\\
	\hline
	$E_1$ (TeV) 	& 	$\Gamma_1$(TeV) 	& 	$E_2$(TeV)		& 	$\Gamma_2$(TeV) 	\\ \hline
	9.101	&	0.09504				&	9.121	&	0.09467	 \\ 
	9.072	&	0.03206				&	9.0923	&	0.03217	 \\ 
	9.003	&$3.417\times10^{-4}$ 	&	9.059	&$3.698\times10^{-3}$	\\ 
	8.965	&$2.734\times10^{-5}$ 	&	9.023	&$3.417\times10^{-4}$	\\ 
	8.925	&$1.956\times10^{-6}$ 	&	8.985	&$2.734\times10^{-5}$	\\ 
	8.884	&$1.274\times10^{-7}$ 	&	8.945	&$1.956\times10^{-6}$\\
	8.842	&$7.641\times10^{-9}$ 	&	8.904	&$1.273\times10^{-7}$\\
	8.798	&$4.258\times10^{-10}$	&	8.862	&$7.641\times10^{-9}$\\
	8.755	&$2.218\times10^{-11}$	&	8.819	&$4.258\times10^{-10}$\\
	$\vdots$&$\vdots$				&$\vdots$	&$\vdots$	\\
\end{tabular}
\end{ruledtabular}
\end{table}

Every potential well has its own resonant energy level and corresponding width. When the incoming energy is near the resonance, the transmission and decay probabilities can be amplified. If the linear term in the potential is turned off, $V(x)$ becomes purely periodic all resonant levels would be the same, thus forming the well-known Bloch band. Because of existence of the tilting linear term, the resonant levels could be different, which would cause the probabilities of transmission and decay to be greatly suppressed when going through more barriers due to the mismatch between resonances at neighboring potential wells. The numerical estimates of resonance levels in first two potential wells are shown in Table~\ref{tab:rlevel}. We see that when the potential is more tilted the mismatch of nearby resonances is more severe, which means that the resonance enhancement effect is more suppressed. In other words, the probability of going through barriers would be exponentially small. This can also be seen through direct numerical estimations.

Numerical order-of-magnitude estimates reveal the situation of going through two and three barriers, in which resonance enhanced probability of transmission and decaying are listed in Table~\ref{tab:estimation}. We see that in both cases, the probability of transmission in the three-barrier case is generally much smaller than that in the two-barrier case. Also, compared to the 3 GeV case, the 20 GeV tilt causes a greater suppression of $|\mathcal{T}|^2$ and $|\mathcal{G}|^2$ in both the two- and three-barrier cases. All numerical results in these tables can only used ad the tool to see that resonant enhancement effect does exist, as we lack exact information about the decay behavior.
\begin{table}
\caption{\label{tab:estimation}Estimation of probabilities.}
\footnotesize
\begin{ruledtabular}
\begin{tabular}{lccccc}
	\multicolumn{6}{c}{$c=3 m_{\tiny W}/\pi$ GeV, $\Delta n=2$}\\
	\hline
	$E_{\text{in}}$(TeV) & $\gamma_{D}$(GeV) & $\alpha$ & $\beta$ & $|\mathcal{T}|^2$& $|\mathcal{G}|^2$\\ \hline
	9.056	&7.192&0.09951&0.5183&	0.08012			&0.4769				\\ 
	9.022	&2.621&0.09963&0.7976&0.01221	&0.2172			\\ 
	8.986	&0.8255&0.09976&0.9339&$1.271\t10^{-3}$	&0.07464\\ 
	8.948	&0.2382&0.09989&0.9811&$1.888\t10^{-5}$	&$3.904\t10^{-3}$\\ 
	8.908	&0.06460&0.1000&0.9950&$7.466\t10^{-6}$	&$5.887\t10^{-3}$\\
	\hline
	\hline
	\multicolumn{6}{c}{$c=20 m_W/\pi$ GeV, $\Delta n=2$}\\
	\hline
	$E_{\text{in}}$(TeV) & $\gamma_{D}$(GeV) & $\alpha$ & $\beta$ & $|\mathcal{T}|^2$& $|\mathcal{G}|^2$\\ \hline
	9.072	&7.192&0.5821&0.5173&	0.06436			&0.7090				\\ 
	9.003	&2.621&0.5832&0.8072&$1.379\t10^{-4}$	&0.04507			\\ 
	8.965	&0.8255&0.5838&0.9366&$6.009\t10^{-6}$	&$7.553\t10^{-3}$\\ 
	8.925	&0.2382&0.5844&0.9818&$5.423\t10^{-7}$	&$2.776\t10^{-3}$\\ 
	8.884	&0.06460&0.5851&0.9952&$1.021\t10^{-8}$	&$2.154\t10^{-4}$\\
	\hline
	\hline
	\multicolumn{6}{c}{$c=3 m_W/\pi$ GeV, $\Delta n=3$}\\
	\hline
	$E_{\text{in}}$(TeV) & \multicolumn{2}{c}{$\gamma_{D}$(GeV)} &\multicolumn{2}{c}{$|\mathcal{T}|^2$}& $|\mathcal{G}|^2$\\ \hline
	9.060	&\multicolumn{2}{c}{7.192}&	\multicolumn{2}{c}{0.02024}		&0.4065				\\ 
	9.054	&\multicolumn{2}{c}{2.621}& \multicolumn{2}{c}{0.06084}		&0.7511			\\ 
	9.022	&\multicolumn{2}{c}{0.8255}&\multicolumn{2}{c}{$5.551\t10^{-3}$}	&0.5454\\ 
	8.989	&\multicolumn{2}{c}{0.2382}&\multicolumn{2}{c}{$1.677\t10^{-4}$}	&$3.497\t10^{-3}$\\ 
	8.986	&\multicolumn{2}{c}{0.06460}&\multicolumn{2}{c}{$1.011\t10^{-3}$}	&0.6364\\
	\hline
	\hline
	\multicolumn{6}{c}{$c=20 m_W/\pi$ GeV, $\Delta n=3$}\\
	\hline
	$E_{\text{in}}$(TeV) & \multicolumn{2}{c}{$\gamma_{D}$(GeV)} &\multicolumn{2}{c}{$|\mathcal{T}|^2$}& $|\mathcal{G}|^2$\\ 
	\hline
	9.059	&\multicolumn{2}{c}{7.192}&	\multicolumn{2}{c}{$9.816\t10^{-5}$}	&0.07456				\\ 
	9.039	&\multicolumn{2}{c}{2.621}&\multicolumn{2}{c}{$1.230\t10^{-5}$}	&0.3894			\\ 
	9.023	&\multicolumn{2}{c}{0.8255}&\multicolumn{2}{c}{$6.491\t10^{-6}$}	&$1.356\t10^{-3}$\\ 
	9.003	&\multicolumn{2}{c}{0.2382}&\multicolumn{2}{c}{$1.008\t10^{-6}$}	&0.3517\\ 
	8.985	&\multicolumn{2}{c}{0.06460}&\multicolumn{2}{c}{$4.511\t10^{-7}$}	&$4.966\t10^{-5}$\\
\end{tabular}
\end{ruledtabular}
\end{table}

From above analysis, we argue that it is possible for incoming wave going through one or two barriers unsuppressed. Overall, for left-incoming wave, we expect that the $\Delta n=1$ process is much more  likely than the $\Delta n=2$ process.

\section{Discussion}
\label{Disc}

Earlier work~\cite{Tye:2015tva} argued why ($B+L$)-violating processes may not be exponentially suppressed for two-particle scattering at energies close to and above the sphaleron energy. However, in view of the QM analysis, it is not clear what the Bloch-wave analysis will lead to: a single sphaleron transition, multisphaleron transition, or no transition. Naive arguments seems to suggest that the no ($B+L$)-violating transition will take place. Here we point out that decay is necessary for the resonant tunneling phenomenon to take place for a sphaleron potential like that shown in Fig.~\ref{p}.

For energies much lower than the sphaleron energy of $9$ TeV, the band widths are too narrow to be relevant, so we focus on incoming energies close to and above the sphaleron energy. At the 14 TeV LHC proton-proton run, the rate of quark-quark scattering with $9$ TeV incoming quark-quark energy is suppressed by the parton distribution function (about $10^{-6}$) and by the phase-space suppression factor of about $10^{-4}$ [probability of $9$ TeV energy in the ($B+L$)-violating direction instead of sharing it with the other ($B+L$)-conserving directions] so the ($B+L$)-violating cross section is suppressed by 9 orders of magnitude just from phase-space considerations. Due to the uncertainties in the estimate given in Ref.~\cite{Tye:2015tva}, the detection of ($B+L$)-violating processes in the 14 TeV run is not assured even if the overall Bloch-wave picture is correct. Increasing the proton-proton energy will go a long way in enhancing the parton distribution probability as well as the available phase space (i.e., $E_{||}$ versus $E_{qq}$) so that the ($B+L$)-violating scattering processes have a much better chance of being observed. 

\vspace{4mm}

\begin{center}
{\bf Acknowledgments}
\end{center}

\vspace{2mm}

We thank Andy Cohen and Sam Wong for many valuable discussions. We thank Ira Wasserman who reminded us that the triple-$\alpha$ process should be considered as a resonant tunneling phenomenon. This work is partly supported by AOE Grant AoE/P-404/18-6 of the Research Grant Council (RGC) of Hong Kong.

\appendix

\section{The presence of continuous bands}
\label{bands}

For the sake of completeness and clarity, here we review the argument that Bloch-wave bands do exist in EW theory. In the absence of left-handed fermions, the periodic sphaleron potential in the $SU(2)$ gauge theory maps to a circle (pendulum), so there are no continuous bands, and the ground state is described by the $\theta$ vacuum $|\theta \rangle$. In the presence of quarks (and leptons), the vacua have different baryon numbers, implying that the sphaleron potential is actually a periodic potential which allows continuous Bloch-wave bands. 

Note that, as is well known, there are no Bloch-wave bands in QCD even though the quarks couple to the gluon fields. This is because the quark currents coupled to the $SU(3)$ gauge fields are vector-like and remain conserved. The corresponding axial currents are anomalous, but they do not carry baryon number. In contrast, the anomalous left-handed fermionic currents in the EW model do carry baryon and lepton numbers.

Our discussion here follows that in Ref.~\cite{Bachas:2016ffl}. Consider a particle moving in a one-dimensional periodic potential with period $\pi$, $V(\mu) = V(\mu+ n\pi)$, $n \in {\bf Z}$. However, the model is not fully specified: the translational symmetry is local (gauged) or global. If it is gauged (local), $\mu$ plays the role of the angle of a circle for a pendulum, with the action
\beq
\label{bta}
S= \int \rmd t \mathcal{L}=\int \rmd t \left[\frac{1}{2}M{\dot{\mu}^2}-V(\mu)-\theta\dot{\mu}/\pi\right]\;,
\eeq
where the topological term is due to a magnetic flux through the circle and plays the role of the $\theta$ term in the $SU(N)$ gauge theory.

In the presence of left-handed fermions coupled to the $SU(2)$ gauge fields, conserved currents in Eq.~\eqref{conservQ} show that $(n-3B)$ is conserved, so we have to introduce the constraint
\beq
\mu/\pi - 3B=\text{const.}
\label{consmuB}
\eeq
(Since $B-L$ is conserved, we simplify the discussion by ignoring the leptons.) Let us introduce a Lagrange multiplier $\lambda$ into $\mathcal{L}$ in Eq.~\eqref{bta} instead of the $\theta$ term, 
\beq
\mathcal{L}'=\frac{1}{2}M{\dot{\mu}^2}-V(\mu)--c|B|-\lambda\left(\dot{\mu}-3 \pi \dot{B}\right)\;,
\label{LmuB}
\eeq
where we also introduce the term $c|B|$ to indicate that the presence of baryon masses lifts the energy of the ground states. So now the (approximate) translational symmetry is global. This yields
\begin{align*}
	M\ddot{\mu}+\frac{\rmd V(\mu)}{\rmd \mu}-\dot{\lambda}&=0\;,\\
	3\pi \dot{\lambda}\pm c&=0\;,\\
	\dot{\mu}-3 \pi \dot{B}&=0\;,
\end{align*}
where $\pm$ depends on $B$ being positive or negative. Choosing $\mu=B=0$ as the starting point, a slight rearrangement gives
\begin{align}
	\label{newmuB}
	M\ddot{\mu}+\frac{\rmd}{\rmd \mu}\left[V(\mu)+\frac{c}{3\pi}|\mu| \right]&=0\;,
\end{align}
yielding the potential \eqref{cV} used earlier. This system can be interpreted as a particle moving in the $(\mu-3\pi B)$-conserving direction, as shown in Fig.~\ref{muBplot}, with a potential $V_{\text{eff}}(\mu)$ breaking the original periodic structure that $V(\mu)$ possesses. Even for $c=0$, we see that the translational symmetry is global, since the baryon number $B$ is different for different $|n \rangle$ states. So continuous Bloch-wave bands are present.

One may add a kinetic term for $B$ inside $\mathcal{L}'$ in Eq.~\eqref{LmuB}. Because of the constraint~\eqref{consmuB}, the $\ddot B$ term merges with the $\ddot \mu$ term in Eq.~\eqref{newmuB}. This will introduce a modification of the mass $M$ here and the mass $m$ in Eq.~\eqref{Bloch}.

\section{The Triple-$\alpha$ Process in Nucleosynthesis}
\label{triple}

The resonantly enhanced tunneling has been applied to the triple-$\alpha$ process to create carbon-12 in stars. Based on the existence of carbon and higher elements in nature, the resonant state was predicted by Salpeter and Hoyle in 1953 and quickly confirmed in experiment~\cite{Salpeter1952,Hoyle1953}. In the late evolutionary stage of stars, the temperature becomes high enough so that helium starts to burn. Through triple-$\alpha$ process, carbon is produced via a resonance enhancement. Without this excited carbon resonance, carbon and higher elements would not be formed.
 
The nuclear reaction we are interested in is the second step of the triple-$\alpha$
process, 
\[
	^4\text{He}+^4\text{He}+ ^4\text{He}\quad \to \quad 
	^4\text{He}+^8\text{Be}\quad \to \quad ^{12}\text{C}^*
\]
where $^{12}\text{C}^*$ stands for the resonance state in carbon-12.
This two-body collision problem can be simplified as one particle with reduced mass $m_{\text{r}}$ going through an effective repulsive Coulombic potential of the beryllium nucleus. Because both the excited and ground state of carbon-12 are $s$-wave and the reaction takes place in a background of helium gas, we can consider only the direct collisions and apply the time-independent Schr\"odinger equation with reduced mass $m_r$ and potential $\textbf{V}(r)$, which is the combination of Coulombic repulsion and nuclear attraction, similar to the potential shown in Fig.~\ref{pic1}. When the helium tunnels through the potential barrier, it forms the excited state $^{12}\text{C}^*$, which has a number of decay channels. There are two important channels: one is decay into the ground state $^{12}\text{C}$, and the other is decay back to $^4$He and $^8$Be. These two channels correspond to the parameters $\Gamma_{a,b}$ in Eq.~\eqref{eq.decay}. Thus, we have the formula for the probability of such a reaction producing the ground state of carbon-12. With the classical collision rate between the nuclei of $^4$He and $^8$Be, we can write the production rate of carbon-12 as
\begin{widetext}
\beq\label{c12p}
		\frac{\rmd n_{12}^*}{\rmd t}=
			n_4 n_8 \pi {r_0}^2 (4^{\frac{1}{3}}+8^{\frac{1}{3}})^2
			\left(\frac{8}{m_{\text{r}} \pi}\right)^{\frac{1}{2}} 
			\left(\frac{1}{k_B T}\right)^{\frac{3}{2}}
			\int_{\text{fusion window}} E\: \exp{\left(-\frac{E}{k_B T}\right)}
			|\mathcal{G}(E)|^2\; \rmd E\;,
\eeq
\end{widetext}
where $n_{4,8}$ are the particle concentrations of helium and beryllium, respectively. If we consider helium burning at a temperature $T=2\times10^8$ K and density $\rho=10^8 \; \text{kg}\cdot\text{m}^{-3}$, then $n_4=1.5\times10^{34}\, \text{m}^{-3}$ and $n_8=7\times10^{26} \;\text{m}^{-3}$. $|\mathcal{G}(E)|^2$ is a function of the incoming energy $E$ as given in Eq.~\eqref{eq.decay}, which describes the probability of an incoming particle staying inside the potential well. The helium obeys the Maxwell-Boltzmann distribution, which drops rapidly as $E$ increases, while tunneling is exponentially suppressed as $E$ decreases. As a result, the helium distribution that penetrates the barrier has a peak (i.e., the Gamow peak)  in energy $E_G$. The position of this peak and its width is the ``fusion window." As predicted, this is where the resonance level of carbon-12 is located. According to Ref.~\cite{FREER20141}, the total decay width of the state is $9.3\; \text{eV}$ and the radiative decay width is $3.7\times10^{-3}\; \text{eV}$. This gives the stellar synthesis rate of carbon-12 $\frac{\rmd n_{12}^*}{\rmd t} \sim 10^{30}\; \text{m}^{-3}\cdot\text{s}^{-1}$. If the $^{12}\text{C}^*$ resonance is absent (or off by a fraction of an MeV), or its decay is slower, the synthesis of carbon will be very much suppressed. This case clearly illustrates the need for the resonance as well as its decay after tunneling.

\section{Sphaleron in Minkowski Spacetime}
\label{Mink}

\begin{figure}
		\centering
		\includegraphics[height=8cm]{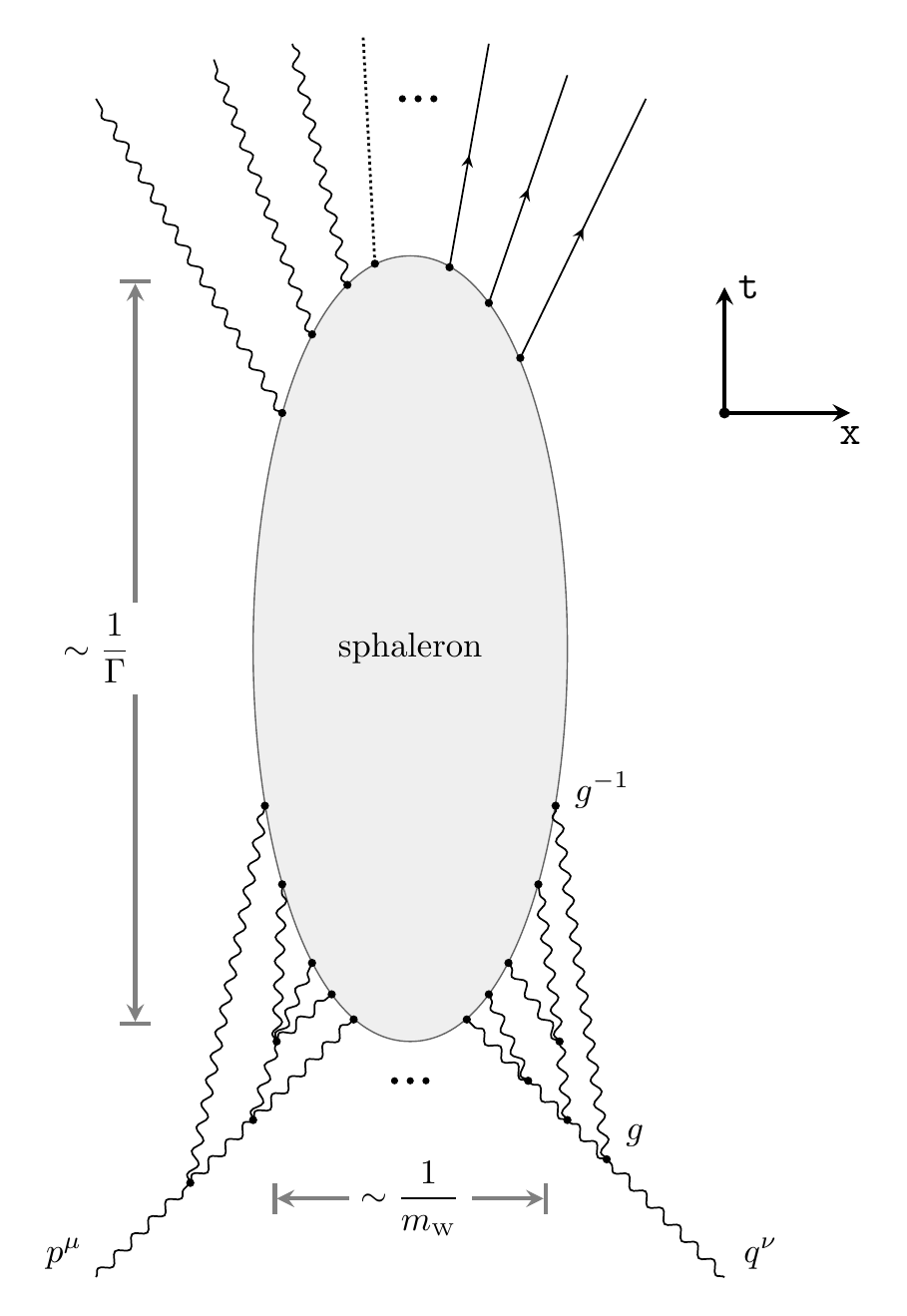}
		\caption{A schematic picture of how two high-energy $W$ bosons can produce a sphaleron in Minkowski spacetime. Each vertex factor of $g$ from their peripheral emission is canceled by a $1/g$ when it hits the sphaleron. So the initial process (with an infinite set of diagrams) has a leading-order contribution of $g^{-2}$. Each final-state $W$ boson has a factor of $1/g$ associated with it; thus, there is the possibility of enhancing the ($B+L$)-violating rate with multiple production of $W$ bosons~\cite{Ringwald:1989ee}.}
		\label{f5}
\end{figure}

Here we take the opportunity to make a few comments on the ``few-to-many" issue for the initial state. An instanton in Euclidean space is four-dimentional and spherically symmetric. allowing all available sizes. On the other hand, in tunneling through a sphaleron barrier in Minkowski spacetime, we only have three-dimensional spherical symmetry, where its size is determined by the $W$ boson mass $m_W$ and the Higgs boson mass $m_H$. Now, it is straightforward to treat the Chern-Simons variable $\mu(t)/\pi$ as a function of Minkowski time~\cite{Tye:2016pxi}. Although the choice of $\mu(t)$ is gauge dependent, it is gauge invariant at $\mu/\pi  \in {\bf Z}/2$  and can be identified with the topological Chern-Simons or Hopf index at $\mu/\pi  \in {\bf Z}$. 

For energies just above or just below the sphaleron energy, we find that the width $\Gamma$ of the Bloch wave is of the order of 10 GeV~\cite{Tye:2015tva}, yielding a time scale of $\delta t \sim \hbar/\Gamma$. This should be compared with the width of the sphaleron in the spatial direction, dictated by $m_W$ and $m_H$ (of the order of 100 GeV). For lower-energy Bloch waves, the $\mu(t)/\pi$ lasts much longer. This is in line with the argument given in Ref.~\cite{Li:1991xq} that is worth repeating here.

For two $W$ bosons with momenta $p^{\mu} = (E, 0, 0, p)$ and $q^{\mu} = (E, 0, 0, -p)$ in high energy ($E\gg m_W$) scattering, we have $p^+ = E + p \sim  2E$ and $q^- \sim 2E$ while $p^- \sim m_W^2/2E$ and $q^+ \sim m_W^2/2E$. In position space, since
\[
x^{\pm} \sim 1/p^{\mp}, \quad \quad  y^{\pm} \sim 1/q^{\mp}
\]
as in multiperipheral scattering, the characteristic distance probed by their scattering is
\[
(x-y)^2 \sim (x^+-y^+)(x^--y^-) \sim -x^+y^- \sim -E^2/m_W^4
\]
which is large. In general, the sphaleron scattering behaves quite different from that of the instanton. This is illustrated in Fig.~\ref{f5}, where $\mu(t)$ is more extended along the time direction than along the spatial directions. The initial scattering at $1/\alpha$ order in the coupling is shown. This includes all tree diagrams in which there is no exchange of bosons. Figure~\ref{f5} suggests that the creation of such a sphaleron is not suppressed by any power in the coupling.

So far, our discussion has ignored the $U(1)$ gauge field in the standard EW theory. Turning it on will lower the sphaleron mass from 9.1 to 9.0 TeV~\cite{Klinkhamer:1984di}. It is interesting to note that the resulting sphaleron may be described as a virtual magnetic monopole-antimonopole pair, as a stable magnetic monopole does not exist in EW theory. 

\bibliography{reference}

\end{document}